\documentclass[prfluids,a4paper]{revtex4-2}
\usepackage[utf8]{inputenc}
\usepackage[T1]{fontenc}
\usepackage{amsmath, bm, commath, mathtools, physics}
\usepackage[colorlinks]{hyperref}
\usepackage[capitalise]{cleveref}
\usepackage{siunitx, booktabs}
\usepackage[table,dvipsnames]{xcolor}

\begin{document}

\title{Effect of electromagnetic boundary conditions on the onset of small-scale dynamos driven by convection}
\author{M. Fontana}
\email{mfontana@df.uba.ar}
\author{P. D. Mininni}
\author{P. Dmitruk}
\affiliation{Universidad de Buenos Aires, Facultad de Ciencias Exactas y Naturales, Departamento de Física, Ciudad Universitaria, 1428 Buenos Aires, Argentina,}
\affiliation{CONICET - Universidad de Buenos Aires, Instituto de F\'{\i}sica del Plasma (INFIP), Ciudad Universitaria, 1428 Buenos Aires, Argentina.}

\begin{abstract}
We present a high-order numerical study of the dependence of the dynamo onset on electromagnetic boundary conditions, in convecting Boussinesq flows forced by temperature gradients. Perfectly conducting boundaries, vacuum, and mixed boundary conditions are considered, treating fields and boundary conditions with close to spectral accuracy. Having one or two conducting boundaries greatly facilitates dynamo action. For the mixed case it is shown that the critical magnetic Reynolds number becomes independent of the Rayleigh number, $Ra$, for sufficiently large $Ra$.
\end{abstract}

\maketitle

\section{Introduction}
Magnetic fields are ubiquitous in nature, defining the dynamics of many physical systems from the atomic to the galactic scale. Of particular interest is the mechanism by which magnetic fields are amplified inside celestial objects, such as the Earth or the Sun, to name two examples. The commonly accepted explanation for said origin is the transformation of kinetic into magnetic energy in an electrically conducting fluid inside the object, i.e., a magnetohydrodynamic (MHD) dynamo \cite{Pouquet1976, Brandenburg2005, Verhille2010, Tobias2021, Mannix2022}. In the Earth the mechanism that drives the flow is thought to be thermochemical convection due to interior cooling \cite{Jones2007, Roberts2013}, while in the Sun convection drives the small-scale dynamo responsible for the magnetic carpet \cite{Kitiashvili2015}.

Although the dynamo mechanism was first proposed at the beginning of the 20th century, it was quickly realized that simple symmetric flows are unable to sustain the magnetic fields observed in nature \cite{Cowling1933, Krause1980}. This notably increases experimental and numerical difficulties in reproducing dynamo generation \cite{Gailitis2000, Muller2002, Nornberg2006, Schaeffer2017, Fontana2018}. Only in the last two decades laboratory experiments and simulations attained dynamos which exhibit features found in celestial magnetic fields, such as a high degree of symmetry, reversals, or periodic modulation \cite{Glatzmaier1995, Monchaux2009, Berhanu2010}. The task of experimentally obtaining a self-sustaining dynamo from flow conditions which more closely resemble those found in celestial bodies is still being actively pursued \cite{Olson2013, Cooper2014, Zimmerman2014, Rojas2021}.

Electromagnetic boundary conditions are known to affect the feasibility of attaining self-sustaining dynamos at fixed power input, together with domain geometry, forcing mechanisms, and fluid transport coefficients \cite{Ponty2005, Schekochihin2005, Jackson2013, Sadek2016, Skoutnev2021, Varela2017}. It was recently shown for rotating convective dynamos \cite{Yan2021} that replacing insulating surroundings with perfect conductors greatly affects magnetic topology \cite{Kolhey2022}. However, numerically studying different boundary conditions on equal footing, with the same numerical convergence, provides a challenge. The best confirmation of the influence of electromagnetic boundary conditions on dynamo feasibility is given by the VKS experiment \cite{Monchaux2009}. In this liquid sodium experiment it was shown that changing propellers from stainless steel to iron resulted in measurable differences on the threshold for magnetic amplification.

In this letter we present a study of the influence of electromagnetic boundary conditions on the minimum magnetic Reynolds number required for dynamo action on convecting flows forced by temperature gradients. Direct numerical simulations (DNSs) of the three-dimensional (3D) non-linear MHD equations are employed, using a method which guarantees quasi-spectral convergence for several types of boundary conditions while enforcing solenoidal magnetic fields to machine precision in non-trivial geometries \cite{Fontana2020, Fontana2022}. Perfectly conducting or vacuum boundary conditions are considered, together with a combination of both.
\section{Governing equations}
We consider a magnetofluid in a rectangular domain with dimensions $L_x \times L_y \times L_z = (2\pi \times 2\pi \times 1)L_0$, where $L_0$ is a unitary length. The domain is periodic in the $\hat{\bm x}$ and $\hat{\bm y}$ directions and has impermeable no-slip isothermic walls at $z/L_0=0$ and $1$. Gravity points in the $-\hat {\bm z}$ direction. The dynamics is described under the Boussinesq approximation by the incompressible MHD equations,
\begin{alignat}{2}
	\label{eq:momentum}
	\dpd{\bm u}{t} + (\bm u \cdot \bm \nabla) \bm u &= - \bm \nabla p + \gamma \theta \hat{\bm z} + (\bm \nabla \times \bm b) \times \bm b + \nu \nabla^2 \bm u,
	\qquad \qquad 
	& \bm \nabla \cdot \bm u & = 0,\\
	\label{eq:induction}
	\dpd{\bm b}{t}  &= \bm \nabla \times (\bm u \times \bm b) + \eta \nabla^2 \bm b,
	\qquad \qquad
	& \bm \nabla \cdot \bm b & = 0,\\
	\label{eq:thermal}
	\dpd{\theta}{t} + (\bm u \cdot \bm \nabla) \theta &= \gamma u_z + \kappa \nabla^2 \theta,
\end{alignat}
with mechanical and thermal boundary conditions $\bm u = \bm 0$, $\theta = 0$ at $z/L_0=0$ and $1$. Here, $\bm u$, $\bm b$, and $\theta$ are the velocity, magnetic, and temperature fluctuations; $\nu$, $\eta$, and $\kappa$ are momentum, magnetic, and temperature diffusivities. The parameter $\gamma$ controls the intensity of the buoyancy, which relates to the temperature difference between the plates $\Delta T$ as $\gamma = \sqrt{\alpha \Delta T g /L_z}$, with $\alpha$ the thermal expansion coefficient and $g$ the gravity. The temperature fluctuation is the difference between the total fluid temperature $T$ and the linear profile of the conductive solution $T_0$, $T' = T - T_0$, which is expressed in velocity units as $\theta = \gamma L_z T' / \Delta T$ \cite{Fontana2020}. The magnetic field is expressed in velocity units as $\bm b = \bm B / \sqrt{\mu \rho}$, with $\bm B$ the magnetic field, $\mu$ the permeability, and $\rho$ the mean fluid density.

A set of parameters to describe \cref{eq:momentum,eq:induction,eq:thermal} is given by the kinematic Reynolds number $Re$, the magnetic Reynolds number $Rm$, the Rayleigh number $Ra$, and the thermal and magnetic Prandtl numbers $Pr$ and $Pm$ \cite{Meneguzzi1989, Jones2000, Fontana2020}. They can be expressed in terms of dimensional quantities as
\begin{equation}
	\def\newexpspacing{4em}
	Re = \frac{U L}{\nu},  \hspace{\newexpspacing} Rm = \frac{U L}{\eta}, \hspace{\newexpspacing} Ra = \frac{\gamma^2 L^4}{\nu \kappa}, \hspace{\newexpspacing} Pr = \frac{\kappa}{\nu}, \hspace{\newexpspacing} Pm = \frac{\eta}{\nu},
\end{equation}
with $U$ and $L$ characteristic speeds and lengthscales. We consider $U$ as the r.m.s.~flow speed, $L=L_z/2$, and $L_z/U$ as the unit of time. $Re$ is the ratio between momentum advection and diffusion. $Rm$ measures the ratio of magnetic induction to dissipation. $Ra$ is the ratio of thermal diffusion time to the time of a fluid element freely ascending due to buoyancy. Finally, $Pr$ and $Pm$ compare the momentum diffusion time against the diffusive time scale of the temperature and magnetic fields.

We consider three scenarios for the electromagnetic boundaries: perfectly conducting walls at $z/L_0=0$ and $1$, vacuum surroundings, and their combination. These conditions are of interest for different situations in geodynamos and stellar dynamos, and for laboratory experiments \cite{Gailitis2002, Monchaux2009, Berhanu2010}. In the first case, perfectly conducting walls approximate the inner-to-outer Earth core boundary \cite{Wicht2002}. In that case there can be no magnetic flux nor tangential electric field at the walls \cite{Jackson1975}. This implies the following relations at either $z/L_0=0$ or $1$ (or at both boundaries):
\begin{equation}
	\bm j \times \hat{\bm z} = \bm 0,
	\qquad \qquad 
	\partial_t \bm b \cdot \hat{\bm z} = 0,
\end{equation}
where $\bm j = \bm \nabla \times \bm b$ is the current density, which at the boundary is proportional to the electric field $\bm E$ if the walls are stationary. Vacuum surroundings approximate the scenario found in stars, and resemble insulating properties of the Earth boundary between the outer core and the mantle \cite{Jones2007}. If the fluid electrical permittivity is close to that of vacuum, they require the continuity of the magnetic field at the walls, together with the continuity of the normal electric field \cite{Jackson1975}. More precisely, the conditions
\begin{equation}
	\label{eq:b-continuity}
	[\bm b] = \bm 0,
	\qquad \qquad
	[E_z] = 0,
\end{equation}
must be obeyed at $z/L_0=0$ or $1$ (or at both places). Here $[\ ]$ denotes the jump at the boundary. As vacuum surrounds the magnetofluid, the magnetic field in the box exterior must be compatible with a potential representation, as there are no currents there. \Cref{eq:b-continuity} requires the enforcement of this condition for $\bm b$ at the walls. Finally, the mixed or combined case corresponds to one boundary condition (e.g., perfect conductor) at $z/L_0=0$, and the other boundary condition (e.g., vacuum surroundings) at $z/L_0=1$. Besides its relevance for the geodynamo, there are experiments which seek to obtain self-sustaining dynamos employing similar conditions \cite{Zimmerman2014,Cooper2014}. 

A self-sustaining dynamo must amplify a seed magnetic field, increasing the magnetic energy of the system $E^b$ as long as the flow is maintained or until non-linear saturation is reached. For systems at finite magnetic Reynolds number, this implies that magnetic induction must overcome Ohmmic losses. To study dynamo feasibility for a given $Rm$ and geometry, it is common practice to consider the flow to be known, and an initial magnetic energy much smaller than the kinetic one, $E^u$, leading to the kinematic dynamo problem. A common strategy to study this problem theoretically is to neglect the feedback of the magnetic field on the velocity, and thus the equation for the magnetic field becomes a linear eigenvalue problem \cite{Moffatt1978, Roberts2000, Roberts2013, Mannix2022}. Correspondingly, to leading order
\begin{equation}
	\label{eq:eb-kinematic}
	E^b (t) \propto e^{\sigma t},
\end{equation}
with $\sigma$ the fastest growing (or slowest decaying) eigenvalue. When $\sigma > 0$ \cref{eq:eb-kinematic} only holds for a finite range of time, after which the decoupling of the induction equation stops being appropriate. The critical magnetic Reynolds number of dynamo onset $Rm^{\textrm{crit}}$ is defined by $\sigma(Rm^\text{crit}) = 0$, and can be determined numerically by solving the MHD equations for a small enough random magnetic seed and for a large enough exploration of parameter space \cite{Ponty2005}.

We consider the feasibility of self-sustaining convective dynamos employing DNSs of the full non-linear \cref{eq:momentum,eq:induction,eq:thermal}. Simulations are performed using the \texttt{SPECTER} parallel pseudo-spectral code \cite{Fontana2020, Fontana2022}, (available at \cite{SPECTER}). Equations are evolved in time using a double-precision second-order Runge-Kutta method. \texttt{SPECTER} employs standard Fourier decompositions along periodic directions, attaining spectral convergence. Non-periodic directions ($\hat{\bm z}$ in our case) utilize a Fourier continuation method with Gram polynomials (FC-Gram) \cite{Lyon2010a,Fontana2020}, which allow high-order representation of boundary conditions without Gibbs degradation, and hence attain spectral accuracy inside the domain and high-order convergence at the boundaries. The absence of magnetic monopoles ($\boldsymbol{\nabla} \cdot \bm{b}=0$) is enforced to machine accuracy by evolving in time the vector potential $\bm a$ in the Coulomb gauge, defined by $\bm \nabla \times \bm a = \bm b$ and $\boldsymbol{\nabla} \cdot \bm{a}=0$. Particularly relevant for this study, \texttt{SPECTER} can implement different boundary conditions while retaining the spectral convergence properties aforementioned \cite{Fontana2022} (as a reference, typical errors are $\langle (\boldsymbol{\nabla} \cdot \bm{b})^2 \rangle < 10^{-30}$, while quadratic errors in magnetic boundary conditions are $\mathcal{O}(10^{-10})$ or smaller; vertical resolutions in the simulations were chosen to maintain small errors). Previous studies of convective dynamos, using other methods or simplified boundary conditions, can be found in \cite{Thelen2000, Cattaneo2003, Chen2015, Proctor2015, Varela2017}. For details of the method used here, or for the full implementation of the electromagnetic boundaries, see \cite{Fontana2022}.

\begin{figure}[t]
	\centering
	\includegraphics[width=.7\textwidth, keepaspectratio=true]{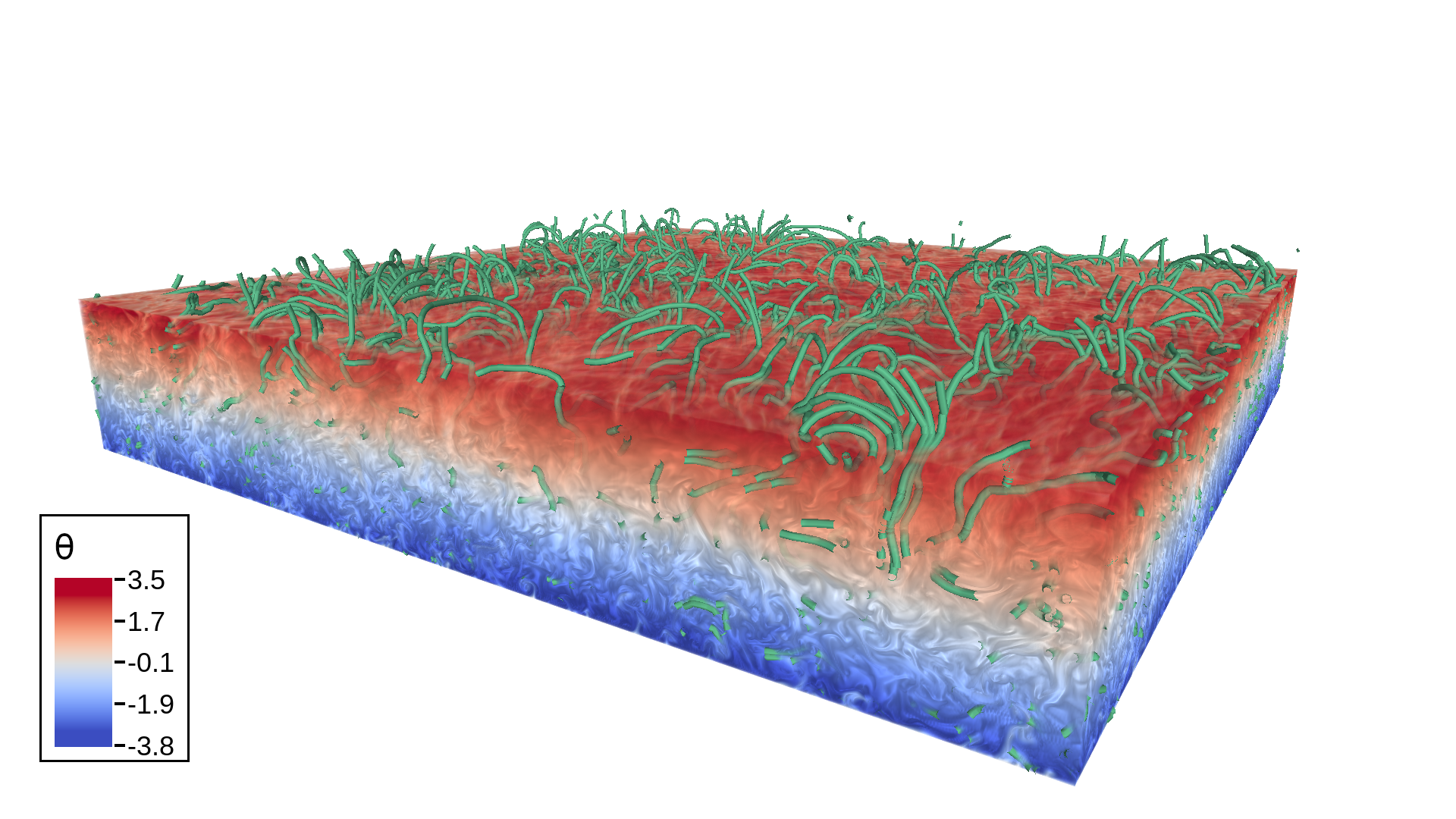}
	\caption{Rendering using the software \texttt{VAPOR} \cite{Clyne2007,VAPOR} of the temperature fluctuations $\theta$ inside the domain, together with the magnetic field lines (in green) for simulation M22. Note that the magnetic field permeates into the vacuum above $z=1$.}
	\label{fig:render}
\end{figure}

\begin{table}
	\begin{center}
		\caption{Simulations with $256 \times 256 \times 103$ grid points and different boundary conditions. Run is the label for each simulation, with C, M, and V denoting conducting, mixed, and vacuum boundaries. $Re$ and $Rm$ are the kinetic and magnetic Reynolds numbers, and $Pm$ is the magnetic Prandtl number. $\sigma$ is the growth rate of the magnetic energy, with negative values corresponding to decaying solutions. All simulations have $Ra=\num{2.25e6}$, $Re=\num{6.06e2}$, and $Pr= \num{1}$.}
		\label{tbl:rm}
		\begin{tabular}{@{\qquad} l @{\qquad} c @{\qquad} c @{\qquad} S[table-format=-1.2] @{\qquad} p{1em} @{\qquad} c @{\qquad} l @{\qquad} c @{\qquad} c @{\qquad} S[table-format=-1.2] @{\qquad}}
			\cmidrule[\heavyrulewidth]{1-4}    \cmidrule[\heavyrulewidth]{6-10}
			\morecmidrules     \cmidrule[\heavyrulewidth]{1-4}    \cmidrule[\heavyrulewidth]{6-10}
			Run& {{$Rm$}}        & {{$Pm$}} & {{$\sigma$}} & & & Run& {{$Rm$}}        & {{$Pm$}}        & {{$\sigma$}}\\
			\cmidrule[\lightrulewidth]{1-4}    \cmidrule[\lightrulewidth]{6-10}
			C1 &$\num{2.43e+02}$ &$\num{4.00e-01}$ & 0.41  & & & M1 &$\num{1.21e+02}$ &$\num{2.00e-01}$ & 0.13\\
			C2 &$\num{1.21e+02}$ &$\num{2.00e-01}$ & 0.32  & & & M2 &$\num{6.06e+01}$ &$\num{1.00e-01}$ & 0.07\\
			C3 &$\num{6.06e+01}$ &$\num{1.00e-01}$ & 0.26  & & & M3 &$\num{2.42e+01}$ &$\num{4.00e-02}$ & -0.01\\
			C4 &$\num{2.41e+01}$ &$\num{4.00e-02}$ & 0.19  & & & M4 &$\num{1.21e+01}$ &$\num{2.00e-02}$ & -0.07\\
			C5 &$\num{1.20e+01}$ &$\num{2.00e-02}$ & 0.11  & & &   &                 &                 &     \\
			C6 &$\num{6.06e+00}$ &$\num{1.00e-02}$ & 0.05  & & & V1 &$\num{6.06e+02}$ &$\num{1.00e+00}$ & 0.40\\
			C7 &$\num{2.40e+00}$ &$\num{4.00e-03}$ & 0.03  & & & V2 &$\num{4.83e+02}$ &$\num{8.00e-01}$ & 0.25\\
			C8 &$\num{1.20e+00}$ &$\num{2.00e-03}$ & -0.00 & & & V3 &$\num{3.03e+02}$ &$\num{5.00e-01}$ & 0.01\\
			C9 &$\num{8.64e-01}$ &$\num{1.43e-03}$ & -0.08 & & & V4 &$\num{2.41e+02}$ &$\num{4.00e-01}$ & -0.07\\
			C10&$\num{6.06e-01}$ &$\num{1.00e-03}$ & -0.32 & & & V5 &$\num{1.21e+02}$ &$\num{2.00e-01}$ & -0.18\\
			\cmidrule[\heavyrulewidth]{1-4}    \cmidrule[\heavyrulewidth]{6-10}
			\morecmidrules    \cmidrule[\heavyrulewidth]{1-4}    \cmidrule[\heavyrulewidth]{6-10}
		\end{tabular}
	\end{center}
\end{table}

\begin{figure}[t]
	\centering
	\includegraphics[width=\textwidth, keepaspectratio=true]{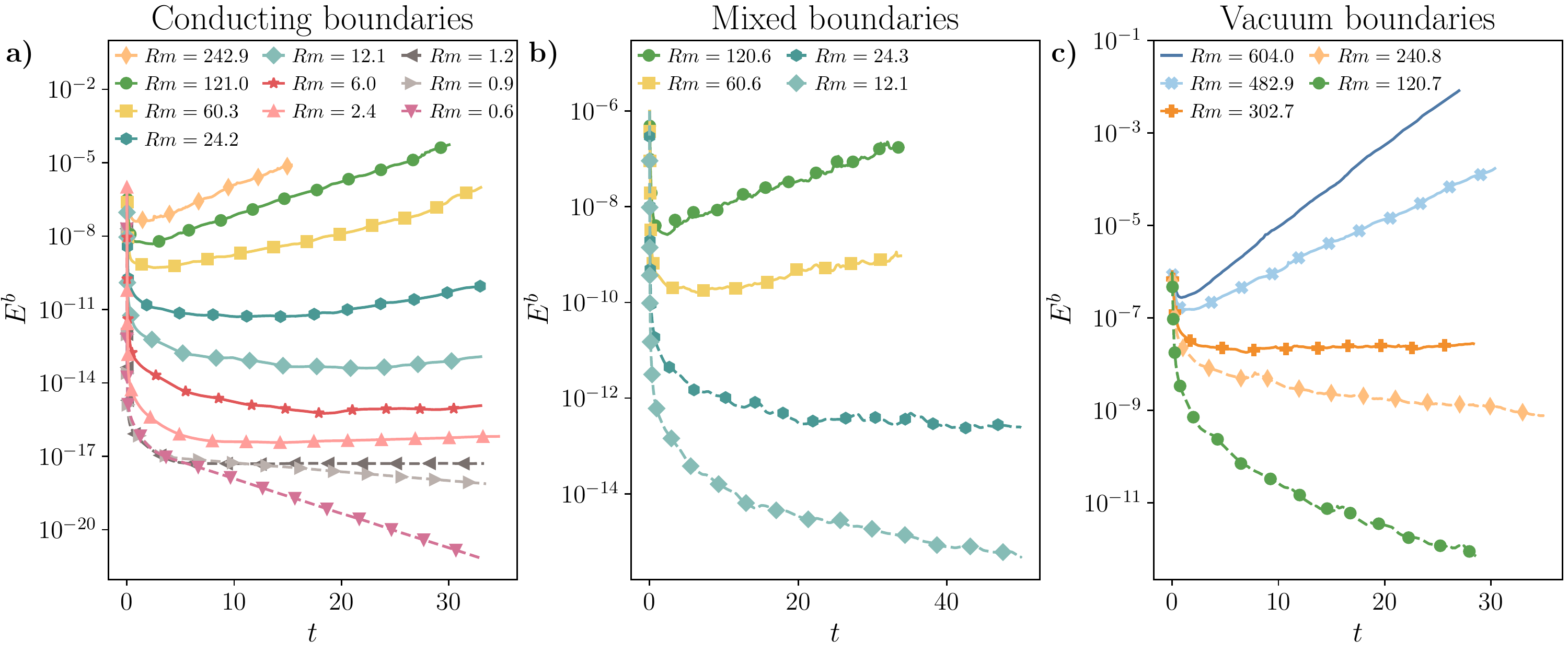}
	\caption{Magnetic energy $E^b$ as a function of time for all simulations at $Ra = \num{2.25e6}$. Line colors and markers denote the magnetic Reynolds number $Rm$ of each simulation. Panels show \textbf{a)} the case of conducting boundaries, \textbf{b)} mixed boundaries, and \textbf{c)} vacuum surroundings. Simulations with decaying magnetic energy are drawn with discontinuous lines.}
	\label{fig:b_vs_t}
\end{figure}
\section{Results}
\subsection{Role of boundary conditions at fixed Rayleigh number.}
We consider first 19 simulations at fixed values of $Ra$ and $Re$, with different magnetic Reynolds numbers and electromagnetic boundaries. We label simulations using $Rm$, even though $Pm$ also changes. As we are interested in the regime in which both velocity and temperature fluctuations are dynamically relevant at all scales, $Pr=1$ is used. A summary of the runs is presented in \cref{tbl:rm}. Also, a rendering of temperature and magnetic field lines for a high resolution run is shown in \cref{fig:render}.

In the simulations we first integrate the non-magnetic Boussinesq equations (\cref{eq:momentum,eq:thermal} with $\bm b = \bm 0$) until a statistically-steady turbulent convective state is attained. The value of $Ra$ is picked to have r.m.s.~flow speeds of order unity, with fields well resolved at the chosen spatial resolution. This state, together with an initial random magnetic field with energy \num{1e-6} (six orders of magnitude smaller than the kinetic energy), is used as initial condition for the MHD simulations in \cref{tbl:rm}. They are integrated in time for each electromagnetic boundary (labeled ``C'' for conducting boundaries, ``M'' for mixed, and ``V'' for vacuum) and value of $Rm$, until a steady exponential growth or decay of the magnetic energy is attained.

In \cref{fig:b_vs_t} the magnetic energy as a function of time is shown for all simulations. An exponential regime is attained in all cases (seen as linear trends in semi-log scale). This is consistent with the fact that a kinematic dynamo should be recovered for small magnetic seeds. Moreover, for all electromagnetic boundaries  both growing and decaying states are present. Note the profound effect of boundary conditions in the growth rate of magnetic energy and, by proxy, in the viability of a self-sustaining dynamo. Decaying solutions (denoted in the figure by discontinuous lines) are present in the vacuum case at values of $Rm$ much larger than those for mixed boundaries or two conducting plates. For the latter case $Rm$ becomes so small in the non-dynamo (decaying) solutions that $E^b$ displays no fluctuations in time, becoming almost purely diffusive.

\begin{figure}[t]
	\centering
	\includegraphics[width=.8\textwidth, keepaspectratio=true]{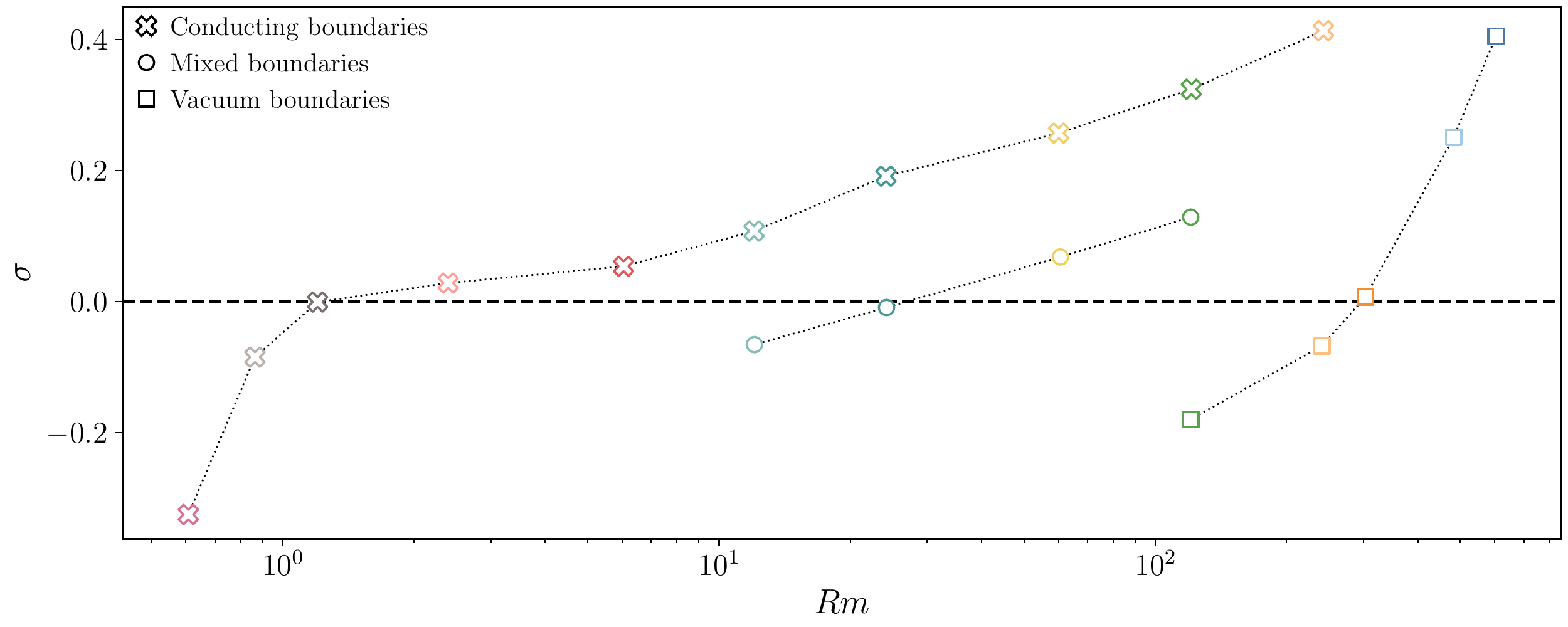}
	\caption{Growth rate $\sigma$ as a function of $Rm$ for simulations with $Ra=\num{2.25e6}$. Markers denote electromagnetic boundary conditions: crosses for conducting boundaries, circles for mixed conditions, and squares for vacuum.}
	\label{fig:growing_rm}
\end{figure}

To characterize the behavior of the growth rate $\sigma$ with $Rm$, we perform a least squares fit of the magnetic energy as a function of time in the range for which an exponential growth or decay can be observed. Results are presented in \cref{fig:growing_rm}, where $\sigma$ is shown as a function of $Rm$ for all simulations. As in \cref{fig:b_vs_t}, there is a major difference for each boundary in the value of $Rm$ for which the zero crossing of $\sigma$ is found. Critical values $Rm^\text{crit}$ to have dynamo onset differ by more than an order of magnitude. For conducting plates $Rm^\text{crit} \approx \num{1.2}$, whereas $Rm^\text{crit} \approx \num{30}$ for mixed boundaries, and $Rm^\text{crit} \approx \num{303}$ for vacuum.

\begin{figure}[t]
	\centering
	\includegraphics[width=\textwidth, keepaspectratio=true]{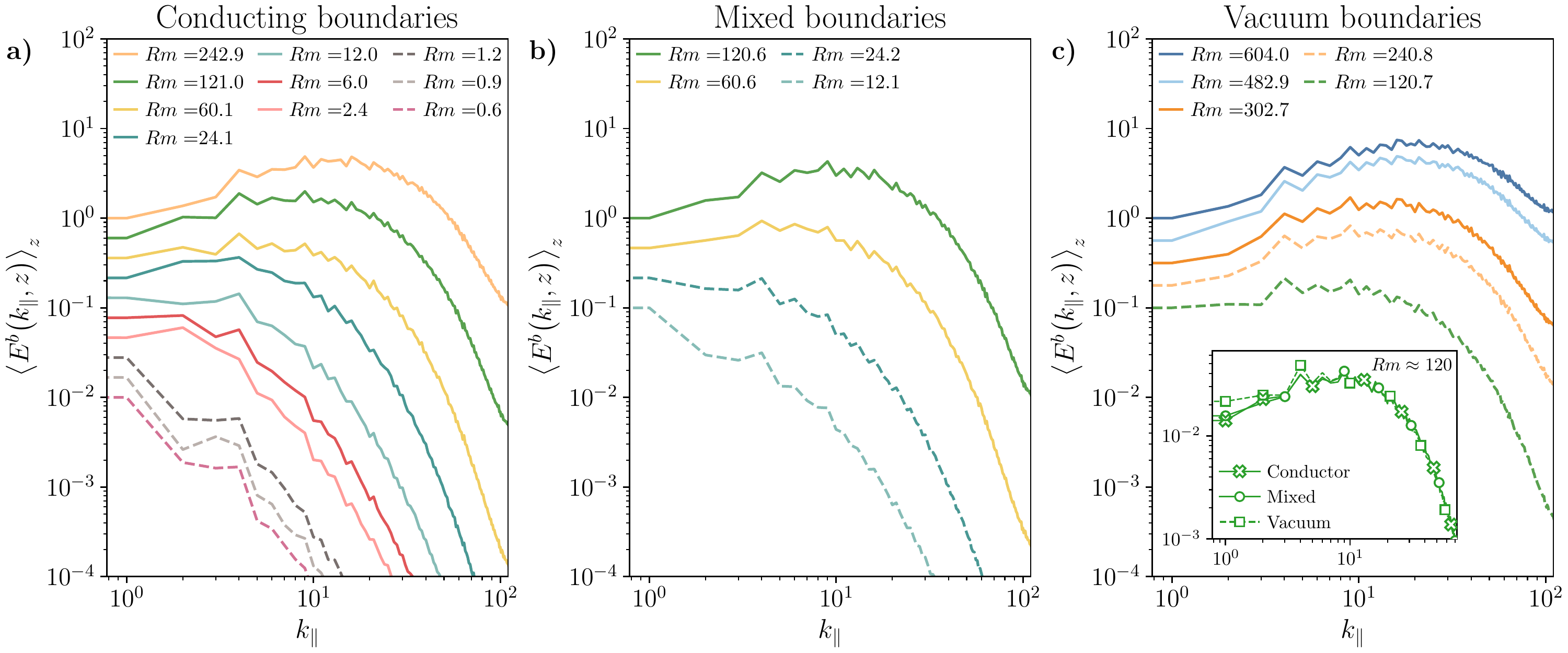}
	\caption{Vertically averaged magnetic energy spectra $\langle E^b(k_\parallel,z) \rangle_z$ for wavenumbers parallel to the wall, $k_\parallel$. \textbf{a)} Magnetic spectra for conducting boundaries. \textbf{b)} Same for mixed, and for \textbf{c)} vacuum boundaries. Decaying solutions are marked with dashes, and colors and symbols reference different values of $Rm$ ($Ra=\num{2.25e6}$ in all cases). The inset in \textbf{c)} shows normalized spectra for three simulations at same $Rm$ differing on boundary conditions (runs C2, M1, and V5). Wavenumbers use $L_x = L_y = 2\pi L_0$ as the reference length. The correlation length of the velocity is similar in $\hat{\bf x}$, $\hat{\bf y}$, and $\hat{\bf z}$, and is $\approx 0.6L_0$ for $u_z$, and $\ell_\parallel \approx 0.75L_0$ (in $\hat{\bf x}$ and $\hat{\bf y}$) and $\ell_z \approx 0.5L_0$ for $u_x$ and $u_y$.}
	\label{fig:spectra_rm}
\end{figure}

To compare the energy-containing scales of the magnetic solutions, we compute the vertically averaged magnetic spectrum as a function of the horizontal wavenumber $k_\parallel = (k_x^2 + k_y^2)^{1/2}$, $\langle E^b(k_\parallel, z)\rangle_z$, and averaged across 10 turnover times. Note \textit{parallel} denotes the directions tangential to the walls, and that the exponential growth or decay is removed before averaging in time by normalizing each spectra by its instantaneous magnetic energy. The result is shown in \cref{fig:spectra_rm}. All self-sustaining dynamo solutions generate magnetic fields with energy concentrated at intermediate or small scales, i.e., the dynamo is a small-scale dynamo. This is expected for flows with negligible helicity \cite{Pouquet1976, Krause1980, Schekochihin2004}. As $Rm$ diminishes (i.e., as Ohmic difussion increases), energy containing structures shift to lower wavenumbers. This can be appreciated in the spectra for all scenarios. For conducting walls, decaying solutions are obtained only when the diffusive scale is approximately the box length, for $Rm \le 1$. This can be seen in \cref{fig:spectra_rm}, where a sharp decrease in $\langle E^b(k_\parallel, z)\rangle_z$ for $k_\parallel = 2$ (i.e., the second smallest wavenumber) characterizes solutions incompatible with self-sustaining dynamos. For mixed and vacuum boundary conditions no such behavior is seen in the spectra for $Rm \approx Rm^\text{crit}$. Finally, in all cases the spectra grow in a self-similar way: all wavenumbers grow with the same rate.

Another interesting observation is found when comparing spectra (normalized by total energy) of the magnetic energy for flows operating at the same $Re$ and $Rm$ but with different boundary conditions. This comparison is shown in the inset in \cref{fig:spectra_rm}c). A clear contrast between scenarios exhibiting self-sustained dynamos (i.e., with conducting or mixed boundaries) and the magnetic extinction case (vacuum) is appreciated. While the three present a considerable accumulation of energy at intermediate scales, the latter has more energy at lower wavenumbers.

\begin{table}
	\begin{center}
		\caption{Simulations with increased values for $Ra$. Run lists the label for each simulation. $N_x \times N_y \times N_z$ indicates the spatial resolution, $Ra$ is the Rayleigh number, $Re$ and $Rm$ are the kinetic and magnetic Reynolds numbers, $Pm$ is the magnetic Prandtl number, and $\sigma$ is the growth rate for the magnetic energy, with negative values corresponding to decaying solutions. All simulations have $Pr= \num{1}$.}
		\label{tbl:ra}
		\begin{tabular}{l c *4{S[table-format=1.2e-1, retain-zero-exponent=true]} S[table-format=-1.2, table-column-width=4em]}
			\toprule
			\toprule
			Run	& $N_x \times N_y \times N_z$ & {{$Ra$}} & {{$Re$}} & {{$Rm$}} & {{$Pm$}} & {{$\sigma$}} \\
			\midrule
			C11 & $512\times512\times231$ 	& 9.00e+06 & 1.16e+03 & 1.16e+01 & 1.00e-02 & 0.009\\
			C12 & $512\times512\times231$ 	& 9.00e+06 & 1.16e+03 & 5.81e+00 & 5.00e-03 & -0.011\\
			C13 & $512\times512\times231$ 	& 9.00e+06 & 1.16e+03 & 2.32e+00 & 2.00e-03 & -0.195\\
			C14 & $512\times512\times231$ 	& 9.00e+06 & 1.16e+03 & 1.16e+00 & 1.00e-03 & -0.353\\
			C21 & $1024\times1024\times487$ & 3.60e+07 & 2.25e+03 & 1.12e+01 & 5.00e-03 & -0.029\\
			M11 & $512\times512\times231$ 	& 9.00e+06 & 1.16e+03 & 2.91e+02 & 2.50e-01 & 0.527\\
			M12 & $512\times512\times231$ 	& 9.00e+06 & 1.15e+03 & 1.15e+02 & 1.00e-01 & 0.012\\
			M13 & $512\times512\times231$ 	& 9.00e+06 & 1.16e+03 & 5.80e+01 & 5.00e-02 & -0.023\\
			M14 & $512\times512\times231$ 	& 9.00e+06 & 1.16e+03 & 2.32e+01 & 2.00e-02 & -0.135\\
			M21 & $1024\times1024\times487$ & 3.60e+07 & 2.25e+03 & 2.81e+02 & 1.25e-01 & 0.329\\
			M22 & $1024\times1024\times487$ & 3.60e+07 & 2.24e+03 & 1.12e+02 & 5.00e-02 & 0.036\\
			M23 & $1024\times1024\times487$ & 3.60e+07 & 2.25e+03 & 5.62e+01 & 2.50e-02 & -0.072\\
			V11 & $512\times512\times231$ 	& 9.00e+06 & 1.14e+03 & 5.68e+02 & 5.00e-01 & 0.029\\
			V12 & $512\times512\times231$ 	& 9.00e+06 & 1.16e+03 & 2.33e+02 & 2.00e-01 & -0.100\\
			V21 & $1024\times1024\times487$ & 3.60e+07 & 2.23e+03 & 5.57e+02 & 2.50e-01 & -0.112\\
			V22 & $1024\times1024\times487$ & 3.60e+07 & 2.25e+03 & 2.25e+02 & 1.00e-01 & -0.212\\
			\bottomrule
			\bottomrule
		\end{tabular}
	\end{center}
\end{table}

\begin{figure}[t]
	\centering
	\includegraphics[width=\textwidth, keepaspectratio=true]{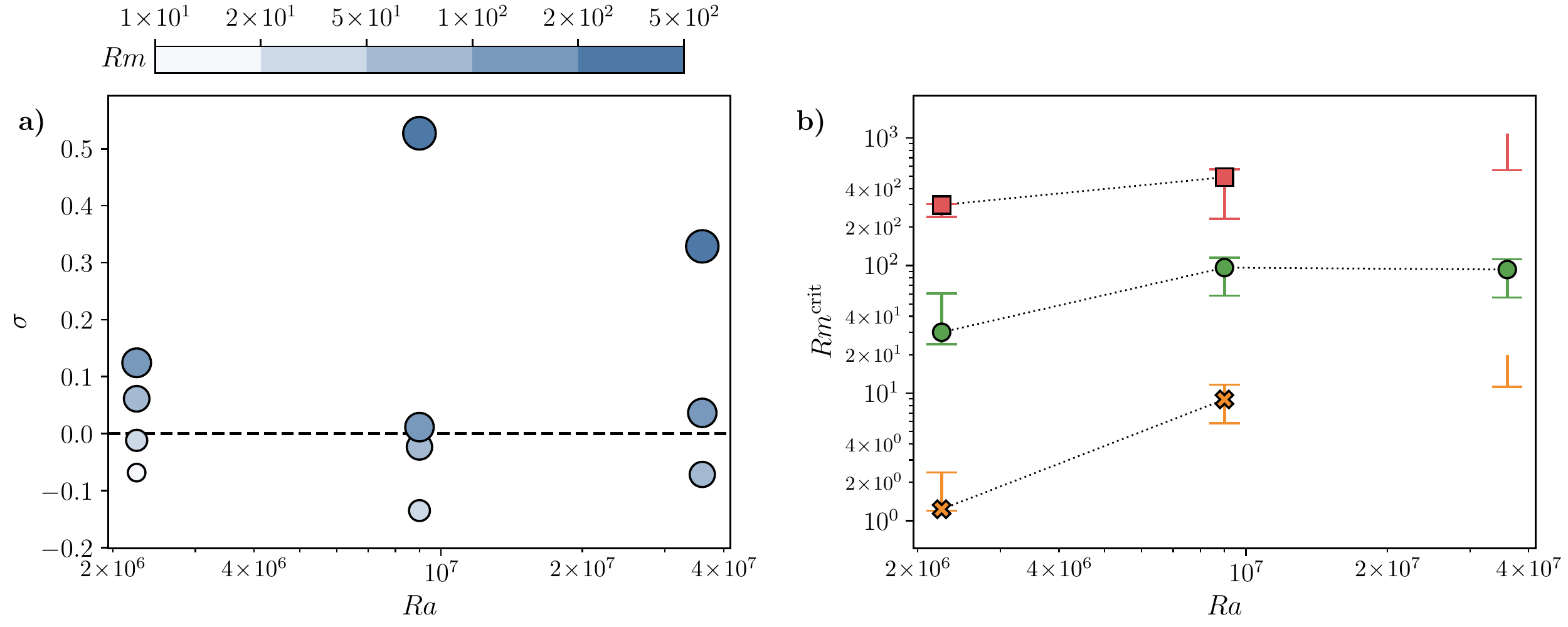}
	\caption{\textbf{a)} Growth rate of the magnetic energy $\sigma$ as a function of the Rayleigh number $Ra$, for mixed boundaries. Increasing values of $Rm$ are denoted with a darker shade and bigger markers. \textbf{b)} Critical magnetic Reynolds number $Rm^\text{crit}$ as a function of $Ra$. Error bars denote the $Rm$ values corresponding to the pair of simulations whose growth rates are closer to zero, and filled symbols correspond to $Rm^\text{crit}$ obtained from a linear interpolation of $\sigma(Rm)$ between those values. Labels for filled symbols are as in \cref{fig:growing_rm}, and open error bars denote lower bounds for $Rm^\text{crit}$.}
	\label{fig:growing_ra}
\end{figure}

\subsection{Role of the Rayleigh number.}

We now study the dependence of $Rm^\textrm{crit}$ on convection strength. To this end we simulate the flow for varying values of $Rm$ as $Ra$ (and hence, $Re$) increases. Note that increasing $Ra$ requires finer grids. \Cref{tbl:ra} lists the simulations with increased values of $Ra$. In all cases the time evolution of the magnetic energy still presents a short transient after which it displays exponential growth or decay. \Cref{fig:growing_ra}a) shows $\sigma$ as a function of $Ra$ and $Rm$. For clarity only the mixed scenario is shown; analogous results are obtained for purely conducting and vacuum boundaries. For fixed $Ra$, lower values of $Rm$ correspond to lower or negative growth rates. More importantly, for fixed $Rm$ the growth rate decreases as $Ra$ increases. This is particularly relevant as naturally occurring dynamos operate at extreme values of $Ra$ (of order \num{1e20} or higher \cite{Olson2007}). Consequently, a determination of asymptotic values of $Rm^\textrm{crit}$ for large $Ra$ would provide better constraints than those in \cref{fig:growing_rm}.

To better analyze this effect we study the behavior of $Rm^\textrm{crit}$ as a function of $Ra$. To obtain a value for $Rm^\text{crit}$ we pick, for each value of $Ra$, the two simulations with growth/decay rates closer to zero. We then estimate a linear relation for $\sigma(Rm)$ between those values and obtain $Rm^\text{crit}$ from its root. The result is shown in \cref{fig:growing_ra}b). $Rm^\text{crit}$ first grows with $Ra$, and for the mixed case it seems to saturate and remain of $\mathcal{O}(10^2)$ even when $Ra$ is increased fourfold. This suggests a finite asymptotic value for $Rm^\text{crit}$ as $Ra \to \infty$, as suggested in \cite{Leorat1981} using other methods. The order of magnitude is compatible with results for the same boundary conditions \cite{Bushby2012}. In the conducting and vacuum scenarios only decaying solutions are available at the highest value of $Ra$ for the values of $Rm$ explored. Thus, only a lower bound for $Rm^\text{crit}$ is marked in the figure. However, simulation C21 has a small $\sigma \lesssim 0$ and $Rm^\text{crit}$ can accordingly be expected to be $\mathcal{O}(10)$, while simulations V21 and V22 have smaller $\sigma$, which suggest $Rm^\text{crit} \sim \mathcal{O}(10^3)$ (see \cref{tbl:ra}). Thus, at least in the range of $Ra$ considered, $Rm^\text{crit}$ differs for the different boundary conditions.

\begin{figure}[t]
	\centering
	\includegraphics[width=\textwidth, keepaspectratio=true]{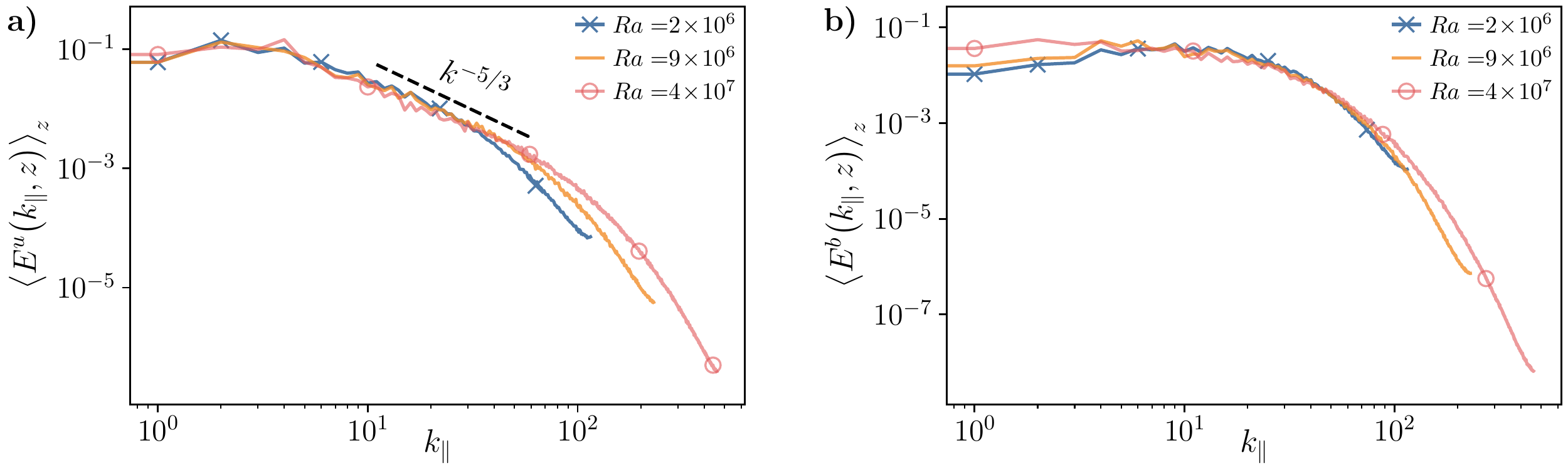}
	\caption{Vertically averaged energy spectra $\langle E(k_\parallel,z) \rangle_z$ as a function of $k_\parallel$ for simulations M4 (blue with crosses), M12 (orange) and M22 (red with empty circles), which operate at approximately the same $Rm$. \textbf{a)} Kinetic energy spectrum $E^u$, and \textbf{b)} magnetic energy spectrum $E^b$. Spectra are normalized by their total energy.}
	\label{fig:spectra_ra}
\end{figure}

Finally, we consider the effect $Ra$ has on the kinetic and magnetic energy spectra. These are shown for runs M4, M12, and M22 (with different $Ra$ but similar $Rm$) in \cref{fig:spectra_ra}. In all cases the spectra are well resolved. In \cref{fig:spectra_ra}a) the kinetic energy content at the largest scales is basically indistinguishable among all simulations. The spectral slope at intermediate scales is similar in all cases, with a power spectrum compatible with $k_\parallel^{-5/3}$ that extends to larger values of $k_\parallel$ as $Ra$ grows. Indeed, it can be readily seen that the kinetic spectrum gets wider as $Ra$ increases. Contrary to the kinetic energy spectra, the width of the magnetic spectra remains the same when $Ra$ increases, as shown in \cref{fig:spectra_ra}b). This is to be expected, as the magnetic Reynolds number is similar for the three simulations. Conversely, the largest scales seem to be more disparate, containing a greater amount of magnetic energy as $Ra$ increases. This indicates that for larger values of $Ra$ magnetic energy grows at slightly larger scales, albeit at a smaller rate.
\section{Discussion}

We studied the problem of attaining dynamos in a channel filled with a conducting fluid undergoing Rayleigh-Bénard convection. To this purpose we performed multiple DNSs of the Boussinesq approximation to the MHD equations, utilizing a versatile numerical method with quasi-spectral convergence \cite{Fontana2022}. We analyzed the growth or decay of an initially small random magnetic field as a function of time, determining the critical magnetic Reynolds number for which self-sustaining dynamo action becomes feasible for three electromagnetic boundary conditions at the floor and the top of the channel and for different values of $Ra$.

Boundary effects are particularly relevant as laboratory experiments have shown strong sensitivity of the dynamo onset to the electromagnetic properties of the containing vessel and the propellers. For the convective case, we analyzed dynamo feasibility for three sets of boundary conditions: both plates perfectly conducting, both walls surrounded by vacuum (with a potential magnetic field outside the domain), and a combination of a conducting floor with vacuum outside the top of the channel. In all cases the magnetic energy was observed to grow or decrease exponentially in time, allowing for the determination of growth or decay rates, and of critical values for dynamo action. Sharp differences, of one order of magnitude or more, were observed in $Rm^\text{crit}$ for each set of boundary conditions, with the fully conducting case showing the mildest constraint and the entirely vacuum scenario displaying the most severe restriction. Consistently with experiments, utilizing conducting materials seems to be notably advantageous for self-sustaining dynamos. Modifying boundary conditions changes the regions of accumulation of magnetic energy: near the walls and with less dissipation for the perfect conductor, and in the bulk with more dissipation for vacuum surroundings.

Raising the strength of convection and turbulence results in a growth of $Rm^\text{crit}$. The observed effect seems to asymptote to values of $Rm^\text{crit}$ independent of $Ra$ for the mixed case, and for all values of $Ra$ explored $Rm^\text{crit}$ differs for the three boundaries considered. In all cases dynamos generate magnetic fields at intermediate scales, with magnetic energy spectra that peak at scales smaller than the flow integral scale.

\begin{acknowledgments}
The authors acknowledge support from CONICET and ANPCyT through PIP, Argentina Grant No.~11220150100324CO, and PICT, Argentina Grant No.~2018-4298. We also thank the Physics Department at the University of Buenos Aires for providing computing time on its Dirac cluster.
\end{acknowledgments}

\bibliography{main}
\end{document}